# Development of a Neuromorphic Network Using BioSFQ Circuits

Evan B. Golden, Vasili K. Semenov, and Sergey K. Tolpygo, *Senior Member, IEEE*

*Abstract*— Superconductor electronics (SCE) appear promising for low energy applications. However, the achieved and projected circuit densities are insufficient for direct competition with CMOS technology. Original algorithms and nontraditional architectures are required for realizing SCE energy advantages for computing. Neuromorphic computing (NMC) is a commonly discussed deviation from conventional CMOS digital solutions. Instead of mimicking a conventional network of artificial neurons, we compose a network from the previously demonstrated single flux quantum (SFQ) electronics components which we termed bioSFQ. We present a design and operation of a new neuromorphic circuit containing a 3x3 array of bioSFQ cells − superconductor artificial neurons − capable of performing various analog functions and based on Josephson junction comparators with complementary outputs. The resultant asynchronous network closely resembles a three-layer perceptron. We also present superconductor analog memory and the memory Read/Write interface implemented with the neural network. The circuits were fabricated in the SFQ5ee process at MIT Lincoln Laboratory.

## I. Introduction

WE are witnessing a skyrocketing growth in demand for artificial intelligence (AI) that has fueled an enormous growth of electric power consumption. Currently, data centers and AI use about 2% of the global energy production, which is projected to double to about 1000 TW·h by 2026, reaching about the energy consumption of Japan. While the largest AI companies develop their own power plants and renewal energy sources, it is not clear if this growth is sustainable.

Most implementations of AI use complementary metal-oxide-semiconductors (CMOS) based hardware and specialized processors. The maturity of CMOS enables cost-effective manufacturing. Implementing neuromorphic circuits with CMOS technology imposes various limitations, e.g.: CPUs, GPUs, and most application-specific digital processors are poorly matched to the inherently analog computations associated with neural networks. AI circuits performance is limited by a low overall efficiency of CMOS where processors are forced to sacrifice performance to prevent overheating. CMOS-based neural networks also impose various architectural constraints. For instance, large heat dissipation requires moving computations to the Cloud, to large server farms. On the other hand, high cost of data transfer and insufficient bandwidth suggests moving AI to Edge computing. These trends are difficult to reconcile. As a result, there is a continuing search for more energy efficient platforms for AI and development of nondigital approaches, e.g., based on analog computing.

One of the alternatives to CMOS-based neuromorphic computing (NMC) platforms, capable of providing comparable complexity while operating faster and with less power dissipation, are networks based on circuits combining CMOS and adjustable two-terminal resistive devices (memristors) and fully memristive neuromorphic networks; see [1], [2] and references therein. Compute-in-memory (CIM) based on resistive random-access memory (RRAM) promises to reduce energy requirements by storing AI model weights in dense, analog and non-volatile RRAM devices, and by performing AI computation directly within RRAM, thus eliminating power-hungry data movement between separate compute and memory; see [3] and references therein.

There have been multiple proposals on using superconductor-based technologies for NMC [4]-[21] because of their energy efficiency even with account for the required cryocooling. Josephson junctions (JJ) act as natural spiking neuron-like devices. Many JJ-based NMC proposals are purely theoretical, and only a few operational JJ-based artificial neural networks (ANNs) have been reported.

In our previous publications [22], [23] we showed that superconductor single flux quantum (SFQ) electronics [24] and its further extensions are inherently suitable for analog computations and can bridge analog and digital computing domains for neuromorphic applications. We demonstrated all

This material is based upon work supported by the Under Secretary of Defense for Research and Engineering under Air Force Contract No. FA8702-15-D-0001. Any opinions, findings, conclusions or recommendations expressed in this material are those of the author(s) and do not necessarily reflect the views of the Under Secretary of Defense for Research and Engineering.

(*Corresponding author: E.B. Golden, e-mail: ev27470@mit.edu*)

E. B. Golden is with the Department of Electrical Engineering and Computer Science, Massachusetts Institute of Technology, Cambridge, MA 02139 USA, and Lincoln Laboratory, Massachusetts Institute of Technology, Lexington, MA 02421, USA (email: ev27470@mit.edu).

S. K. Tolpygo is with the Lincoln Laboratory, Massachusetts Institute of Technology, Lexington, MA 02421, USA (e-mail: Sergey.Tolpygo@ll.mit.edu).

V. K. Semenov is with the Department of Physics and Astronomy, Stony Brook University, Stony Brook, NY 11794-3800, USA (e-mail: Vasili.Semenov@StonyBrook.edu).

The authors on the byline are listed in alphabetical order.

Color versions of one or more of the figures in this article are available online at http://ieeexplore.ieee.org

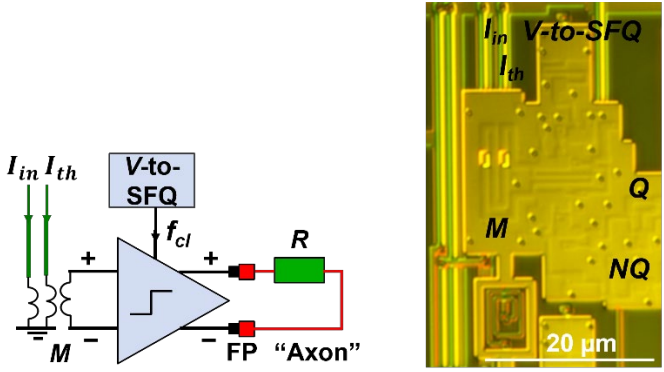
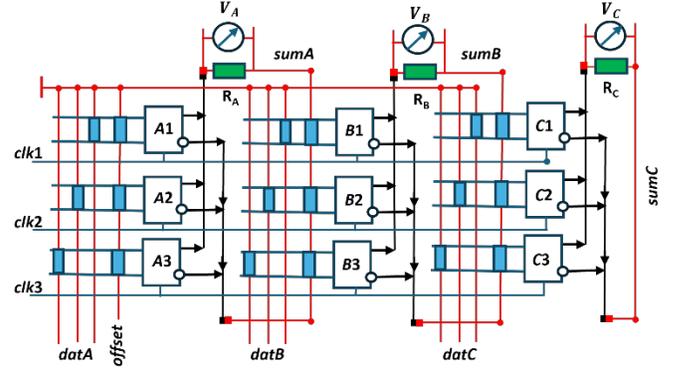

**Fig. 1.** A unit tile used to design a scalable ANN (left panel) and an optical image of the fabricated tile in the circuit (right). The Josephson bipolar comparator has two complementary outputs, $Q$, marked as "+" and NOT($Q$), $NQ$ marked as "−". The comparator's inputs are inductively coupled to PTLs delivering input current signals and the threshold current $I_{th}$ adjustment (offset) via the transformer, $M$. The comparator's outputs pump single flux quanta (SFQ) into a superconducting stripline interrupted by resistor $R$, an analog of the neuron's axon, from its opposite sides, using flux pumps (FP). Local clock frequency $f_{cl}$ is generated by a voltage-biased resistively shunted Josephson junctions, a *V*-to-SFQ converter. A schematic of the circuit with parameter values can be found in [23].

**Fig 2.** Block-diagram of the demonstration circuit consisting of a 3×3 network of superconductor artificial neurons. Blocks Ai, Bi, and Ci are bipolar comparators shown in Fig. 1. Each column - network layer - has a common "axon" PTL interrupted by resistor $R$ shown by a green box.

necessary components of superconducting artificial neurons and demonstrated various operations using analog data encoding such as copying, addition and subtraction, bipolar multiplication, division, square root function, and multiply-accumulate (MAC) function using a bipolar Josephson comparator [22], [23], superconducting inductors (passive transmission lines), and resistors.

In this work, we extended our approach [22], [23] to design, fabricate and demonstrate scalable superconductor ANNs for NMC. For this, we designed a scalable tile for the ANNs and used it to design a new neuromorphic superconductor circuit consisting of a 3×3 array of superconductor artificial neurons and in many respects resembling a three-layer perceptron. The circuit design is presented in Sec. II. The circuit testing results are given in Sec. III. In Sec. IV we present the further development of this circuit, including local memory for storing bipolar analog numbers which network can use as weights and for realization of the back propagation algorithm.

The circuits were fabricated in the SFQ5ee process [25] at MIT Lincoln Laboratory. The circuits were tested in a liquid helium immersion probe using an automated test setup Octopux operating at 1 MHz [26], [27].

## II. SUPERCONDUCTOR BIOSFQ NETWORK DESIGN

### A. The Unit Tile for Scalable Networks

To extend our approach to larger neuromorphic circuits, we developed a demonstration network shown in Figs. 1 and 2. The circuit is composed of the cells reported earlier in [22], [23]. The main component of the ANN unit cell (tile) shown in Fig. 1 is a bipolar Josephson comparator [23] coupled inductively to several passive stripline-type transmission lines (PTLs) serving for delivering the input analog current signals, $I_{in}$ and setting the comparator's local clock frequency $f_{cl}$, using a voltage-to-frequency conversion in a resistively shunted Josephson junction - the Josephson relation $f = V/\Phi_0$; $\Phi_0$ is the flux quantum [24].

The basic operation of the bipolar comparator was explained in [23]. Briefly, it converts the input current $I_{in}$ into two streams of the SFQ pulses at its $Q$ and NOT($Q$), $NQ$ outputs with frequencies given by the probability functions:

$$f_Q = f_{cl}\frac{1}{2}[1 + \mathrm{erf}\left(\pi^{1/2}\frac{I_{in}-I_{th}}{\Delta I}\right)], \quad (1a)$$

$$f_{NQ} = f_{cl}\frac{1}{2}[1 - \mathrm{erf}\left(\pi^{1/2}\frac{I_{in}-I_{th}}{\Delta I}\right)]. \quad (1b)$$

Here, $\Delta I$ is the width of the comparator's grey zone [28], a hardware design-adjustable parameter and $I_{th}$ is the comparator's threshold current that can be adjusted using a control line marked $I_{th}$ in Fig. 1 and *offset* in Fig. 2. In the linear regime of the comparator operation, $\frac{I_{in}-I_{th}}{\Delta I} \ll 1$, these frequencies are simple linear functions of the input current:

$$f_Q = \frac{1}{2}(1 + \frac{I_{in}-I_{th}}{\Delta I})f_{cl}, \quad (2a)$$

$$f_{NQ} = \frac{1}{2}(1 - \frac{I_{in}-I_{th}}{\Delta I})f_{cl}. \quad (2b)$$

Although there is no one-to-one analogy with biological neurons, the bipolar comparator is the nonlinear processing element performing the functions of a neuron soma whereas superconducting PTLs serve as its dendrites. The axon of the superconductor neuron presents a transmission line, a stripline, interrupted by a resistor $R$ [22], [23]; the latter is shown as a green box in Figs. 1 and 2, while the former is a red line connected to the resistor from both sides. The two streams of the SFQ pulses produced by the comparator are pumped from the opposite sides of the transmission line using flux pumps shown by red/black rectangles in Figs. 1 and 2.

The current generated in the axon PTL by the SFQ pulses is proportional to the rate of magnetic flux change in the axon loop, i.e., to $f_Q - f_{NQ}$, and is given by

$$I = \frac{\Phi_0 f_{cl}}{R}\mathrm{erf}\left(\pi^{1/2}\frac{I_{in}-I_{th}}{\Delta I}\right). \quad (3a)$$

In the linear regime of the comparator, the axon current is

$$I = \frac{\Phi_0 f_{in}}{R} \cdot \frac{(I_{in}-I_{th})}{\Delta I}, \quad (3b)$$

that is the weighted and biased input current, $I = wI_{in} + b$, with the positive clock-adjustable weight $w = \Phi_0 f_{cl}/(R\Delta I)$ and adjustable bias $b = -\Phi_0 f_{cl} I_{th}/(R\Delta I)$.

*B. The 3×3 Network*

To construct the first layer of the ANN, we used a column of three identical unit tiles described above and shown in Fig. 2 as units $A1$, $A2$, and $A3$. Each comparator in the units has its own input data PTL. This group of PTLs is marked "*datA*" in Fig. 2. The second, $Bi$, and the third, $Ci$, columns of the network in Fig. 2, the network layers, were constructed similarly to the first column and using the same unit tile; $i$=1,2,3.

The comparators on the network communicate using three types of connections. Red- and blue-colored connections in Fig. 2 represent superconducting PTLs (wires), where the information is coded by the values of flowing currents. Two-color notation is used to present crossing but not connecting wires. Black-colored connections represent Josephson Transmission Lines (JTLs).

Blue-colored wires are also used to deliver "clock voltages" that are converted into the clock frequencies $f_{cl}$ directly in the comparator units. For simplicity and because of a limited number of cryogenic I/O wires to the chip, the comparators in the same row use the same clock voltage (the same clock frequency), marked $clk1$, $clk2$, and $clk3$ in Fig. 2.

Conventional dc voltage bias required for comparators and JTLs is not shown in Fig. 2.

External currents from three data banks, *datA*, *datB*, and *datC*, are applied to the comparators in the first layer, $Ai$ via stripline transformers [29] shown in the Fig. 2 as blue rectangles. These transformers are shown explicitly in Fig. l.

The result of comparing the input current with the threshold currents in each comparator is presented as complementary trains of SFQ pulses. The required input clock pulses are generated by Josephson junctions built into each comparator, i.e., each comparator in the column has its own externally adjustable clock frequency. This has been done to demonstrate that the network is an asynchronous device that can function equally well when its components operate at different or identical clock frequencies. The lack of synchronization is beneficial for mergers of SFQ pulses because it reduces the probability of two coincident pulses arriving at the merger. The SFQ pulse merging in Fig. 2 takes place at all points where any two black lines (black arrowheads) meet and the third line goes out.

Flux pumps, shown as red/black rectangles in Fig. 1 and Fig. 2, pump the merged trains of SFQ pulses from all $Q$ outputs of the comparators in the layer into one side of the axon PTL and from all $NQ$ output into the opposite side of the axon PTL shared by all the comparators in the column (network layer). The flux pumps are nothing more than specialized short JTLs with a wider range of their bias currents, an essential requirement because a fraction of bias current is diverted to the load resistor $R$.

The axon current flowing through the resistor $R_A$ in the first column serves as the input current for the next column (network layer) of the comparators, and so on. The average voltage drops across the resistors $R$ in each layer, $V_A$, $V_B$, and $V_C$, are the main

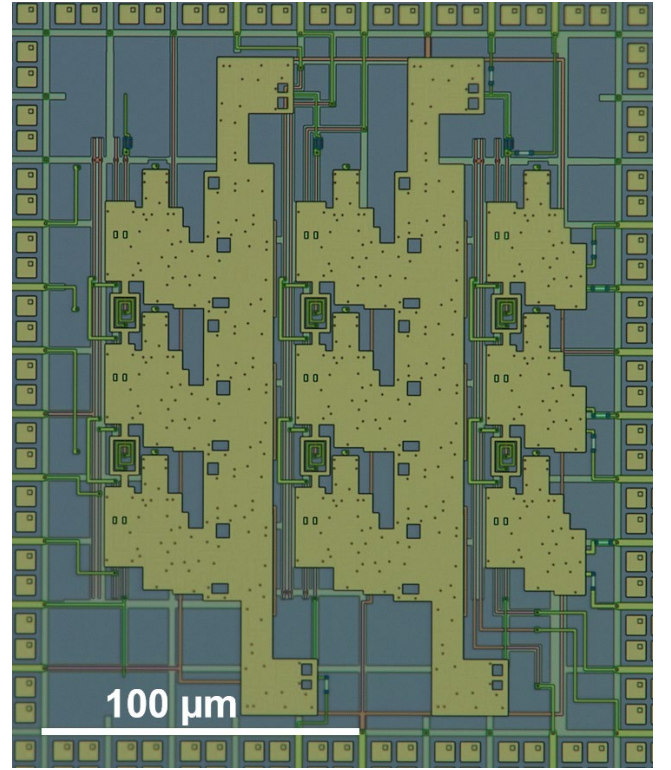

**Fig 3.** Optical image of the fabricated network consisting of a 3×3 array of comparators shown in Fig. 2. Each column of the comparators in the network is clearly visible. The size of each comparator and accumulator unit cell is approximately 20 μm by 50 μm.

output signals from the circuit and measured by room temperature electronics.

The circuit occupies only a tiny fraction of the available space on the 5 mm × 5 mm chip. The density and yield of the used fabrication technology allows for much higher integration levels. The presented network integration scale, the total number of comparators, was limited by the number of I/O pads available in the test probe due to our desire to apply currents to and measure voltages in a larger number of internal points of this demonstration circuit for its potential debugging and detailed investigation.

An optical image of the fabricated network is shown in Fig. 3.

*C. MAC and the Principle of the Network Operation*

Each of the comparators in the first column $Ai$ of the network converts its input current $I_i^A$ into two streams of SFQ pulses, which are applied from the opposite sides of the superconducting transmission line interrupted by the resistor $R_A$ and produce a current given by (3).

In the most general case, the total current induced by the column $A$ comparators in the column's axon inductor and resistor $R_A$ is

$$I^{[A]} = \frac{\Phi_0}{R_A}\sum_i f_{cl,i}^A \operatorname{erf}\left(\pi^{\frac{1}{2}}\frac{I_i^A - I_{th,i}^A}{\Delta I_i^A}\right). \qquad (4a)$$

In the linear regime of all the comparators, $I_i^A - I_{th,i}^A \ll \Delta I_i^A$, (4a) reduces to

$$I^{[A]} = \frac{\Phi_0}{R_A} \sum_i \frac{f_{cl,i}^A}{\Delta I_i^A}(I_i^A - I_{th,i}^A). \quad (4b)$$

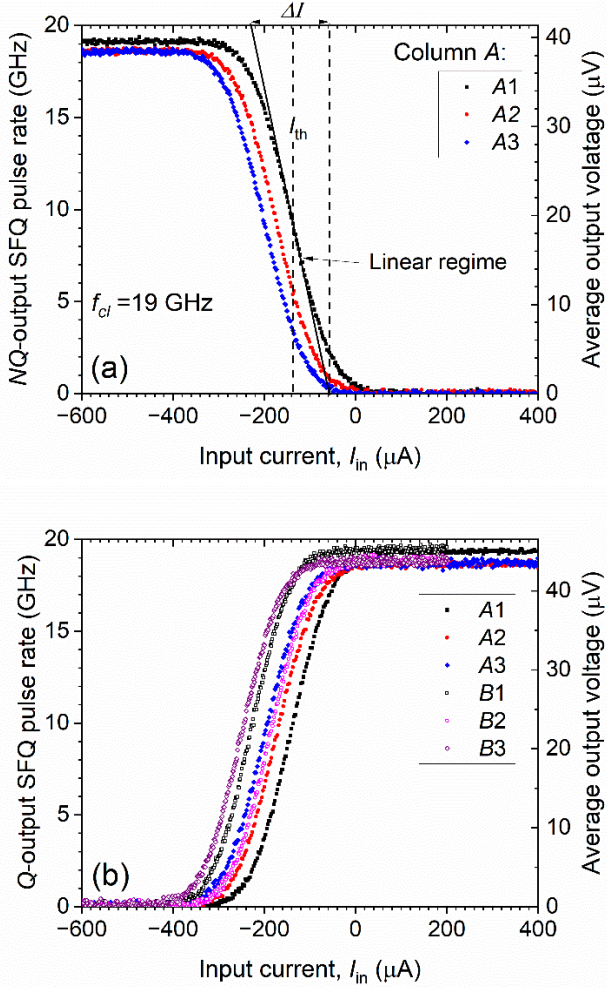

**Fig 4.** Characterization of the individual comparators (artificial neurons) on the 3×3 network in Fig.2: (a) *NQ* output of the comparators in the column *A*. The average output voltage is measured across the resistor $R_A$. In the steady state, this voltage $V_A$ is equal to the rate of SFQ flux change in the axon transmission line due to SFQ pulses coming from the comparators and equal to the product $R_A$ and (4a). (b) *Q*-output of the comparators in the columns *A* and *B*. The average output voltage $V_B$ is measured across the resistor $R_B$. Each comparator was measured individually by disabling all other comparators on the network using the procedure described in the text. Minor variations between the comparators can be observed, and are characterized during the calibration procedure.

That is, the first network layer *Ai* performs a multiply-accumulate (MAC) operation on the input data currents to the first layer of the comparators; the multiplication coefficients (weights) are set by the local clock frequencies and the comparators' grey zone widths $\Delta I_i^A$.

The accumulated current $I^{[A]}$ is then applied to all the comparators in the next network layer, column *Bi*, along with a set of input currents *datB* applied to the individual comparators. Operating in the described manner, comparators in the column-*B* induce current in the column-*B* axon and its resistor $R_B$

$$I^{[B]} = \frac{\Phi_0}{R_B} \sum_i f_{cl,i}^B \, \text{erf} \left( \pi^{\frac{1}{2}} \frac{I^{[A]} + I_i^B - I_{th,i}^B}{\Delta I_i^B} \right). \quad (5a)$$

In the linear regime of the comparators, this current is

$$I^{[B]} = \frac{\Phi_0}{R_B} \sum_j \frac{f_{cl,j}^B}{\Delta I_j^B}\left[\frac{\Phi_0}{R_A} \sum_i \frac{f_{cl,i}^A}{\Delta I_i^A}\left(I_i^A - I_{th,i}^A\right) + I_j^B - I_{th,j}^B\right]. \quad (5b)$$

The accumulated current $I^{[B]}$ is coupled to the inputs of all comparators in the next network layer, column-*C*, along with the set of input currents $I_i^C$ coupled to the individual comparators. In response, the third layer of the network induces the output current in the column-*C* axon and resistor $R_C$

$$I^{[C]} = \frac{\Phi_0}{R_C} \sum_i f_{cl,i}^C \, \text{erf} \left( \pi^{\frac{1}{2}} \frac{I^{[B]} + I_i^C - I_{th,i}^C}{\Delta I_i^C} \right). \quad (6a)$$

In the linear regime of the comparators, this current is

$$I^{[C]} = \frac{\Phi_0}{R_C} \sum_k \frac{f_{cl,k}^C}{\Delta I_k^C} \left\{ \frac{\Phi_0}{R_B} \sum_j \frac{f_{cl,j}^B}{\Delta I_j^B}\left[\frac{\Phi_0}{R_A} \sum_i \frac{f_{cl,i}^A}{\Delta I_i^A}\left(I_i^A - I_{th,i}^A\right) + I_j^B - I_{th,j}^B\right] + I_k^C - I_{th,k}^C \right\}. \quad (6b)$$

To summarize this section, in the linear regime of all the network comparators, the circuit performs linear algebra on its input data sets *datA*, *datB*, and *datC*, using the software-adjustable weights and biases. However, if the excitation of any artificial neuron exceeds the critical level, its response become highly nonlinear resembling the behavior of the biological neurons. This enables the use of the described network for realizing various NMC algorithms, e.g., threshold detection for image recognition. In many respects, the network operates as a three-layer perceptron: column *A* is the input layer; column *B* is the analog of the hidden layer; and column *C* is the output layer.

### III. THE NETWORK CHARACTERIZATION

At the first glance, characterization of the individual comparators in the network is a challenging task. However, we found a relatively simple solution. To characterize comparators, e.g., in the column-*A*, we can measure firstly only a direct, *Q* or complementary, *NQ* cumulative outputs (marked by black color in Fig. 1 and Fig. 2) by measuring the voltage drops, $V_A$, between accordingly left or right terminals of the resistors and the circuit ground. The next step is to disable two of the three comparators by applying to their inputs either large positive or large negative control currents $|I_i - I_{th,i}| \gg \Delta I_i$. In this case, the comparators generate SFQ pulses with 100% probability either on their direct output, if $I_i - I_{th,i} \gg \Delta I_i$, or complementary output, if $I_i - I_{th,i} \ll \Delta I_i$, since $\text{erf}(x) = 1$ at $|x| \gg 1$. This current application disables (passivates) either the direct or complementary outputs of the corresponding comparators allowing testing of the remaining, not disabled, comparator. Another option is to set the clock frequency (voltage on the V-to-SFQ converter) for two rows of the comparators, e.g., with $i$ =2 and 3, to zero, and test sequentially the comparators *A*1, *B*1, and *C*1 in the active row $i$ =1. The latter is achieved by measuring the respective average voltages $V_A$, $V_B$, and $V_C$ which are related by the Josephson relation to the rate of SFQ pulses pumped by the outputs of the corresponding comparators.

Fig. 4 shows an example of such a characterization for several comparators in the network at one particular clock frequency $f_{cl} \approx$ 19 GHz. The test results for the remaining

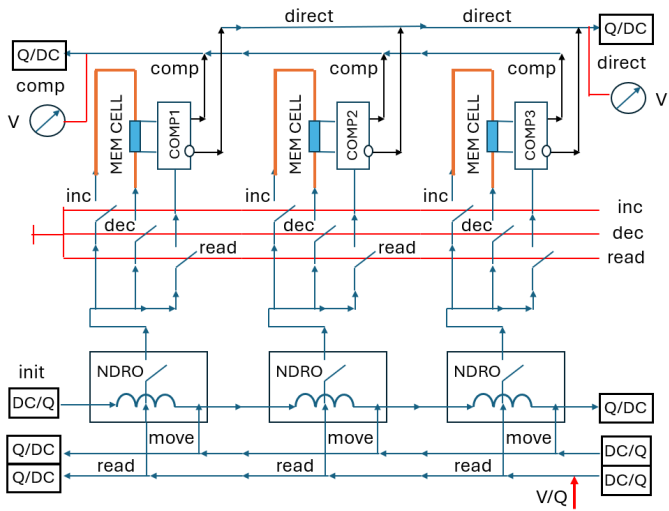

**Fig 5.** Block-diagram of the analog memory for the superconductor neuromorphic circuits. The memory cell inductor shown by the orange line is coupled to a comparator, shown as COMP, by a transformer shown as a blue box. The nine-cell memory has been integrated with the 3x3 network described in Sec. II, and the resultant integrated circuit has been fabricated. Its testing results will be presented elsewhere.

comparators and different clock frequencies look similarly and not presented here to save space.

### IV. ANALOG MEMORY FOR THE NETWORK

We presented the details of our approach to building ANNs for NMC with superconducting SFQ-based components termed bioSFQ [22]. However, some essential components are still missing, e.g., memory. Analog memory is required, for example, for storing weights and implementation of back propagation algorithms. We have developed a memory circuit shown in Fig. 5. The circuit has been fabricated and is under testing.

The main idea behind the memory cell is to remove resistor $R$ shown in Fig. 1. This revision creates the required superconducting memory loop that can store the desired current value. A fragment of the circuit showing three analog memory cells, marked MEM CELL, is presented in Fig. 5. Additional circuitry shown in Fig. 5 is needed to select the memory cell for writing, write information, and read out the stored current value.

The selection of a particular cell is provided by a register of RSFQ NDRO cells located in the lower part of Fig. 5. The initialization of the leftmost NDRO cell is provided by an injection of an SFQ pulse via the DC to SFQ converter (DC/Q) marked as *init*. The activation of the middle or the rightmost cell is provided by applying one or two SFQ pulses via DC/Q converter located in the middle of the right bottom corner of Fig. 5 and connected to line "*move*". After these actions, only one NDRO cell is in a state "1", i.e., stores a fluxon, while two other cells are inactive, i.e., store no fluxons.

To access the memory cell, we apply an SFQ pulse via DC/Q converter located in the lower right corner in Fig. 5 and marked as *read*. The pulse passes only via the active NDRO cell and splits into three JTL lines that are connected with the corresponding normally open switches. The switches can be closed by applying the corresponding dc currents. At the normal operation, only one of three switches should be closed. The *inc* and *dec* switches in Fig. 5 allow for injecting SFQ pulses (fluxons) into the left and right arms of the memory cell inductor coupled to the comparator and, hence, increasing or decreasing the current circulating the superconducting loop.

Reading the memory cell content is performed by the comparator discussed in Sec. 2A and our previous papers [22], [23]. The comparator is clocked by SFQ pulses passed via the closed *read* switch. The direct and complementary output SFQ pulses travel via JTLs and mergers to Q/DC converters shown at the top of Fig. 5.

Up until now, we described how the memory circuit processes single SFQ pulses. The actual circuit has a special voltage-to-SFQ (V/Q) converter, shown by the red arrow near the right bottom corner in Fig. 5, that converts the applied DC voltage to a train of SFQ pulses. When the DC voltage is applied, it is possible to read out direct and complementary outputs of the comparators as the corresponding average voltages, using voltmeters shown near the left and right top corners in Fig. 5.

### V. CONCLUSIONS AND FUTURE WORK

We presented the circuit design and operation of the 3×3 bioSFQ network analogous to a three-layer perceptron, the bioSFQ memory, and the memory Read/Write interface. The complete results of their testing will be presented elsewhere. The network can be scaled up using the developed unit tile, and it would be incorrect to stop at the achieved level of integration. Ultimately, we would like to create a Neuromorphic Processing Unit (NPU) that could coexist with CPUs, GPUs and other well-known groups of processing units.

As an additional illustration of the unique features of our neuromorphic technology we draw your attention to the idea that the circuit in Fig. 5 can also perform arithmetic calculations. In the memory description in Sec. IV, we requested only one NDRO cell to be in the state "1". Now we change this requirement and, using *init* and *move* pulses, program all NDRO cells to be in the state "1". As a result, the *read* pulses activate all three comparators, and their output signals represent direct and complementary values of the sum of the contents of all three memory cells.

The convenience of our bioSFQ approach is that all available RSFQ cells and solutions can be readily implemented, and the new functionality is basically achieved by "rewiring" the known components. Many other possibilities open up by changes in their functionality. For instance, the resistor $R$ interrupting the axon's PTL in Figs. 1 and 2 can be replaced by a Josephson junction. In this case, the axon loop can locally store currents lower than the critical current of the junction. This creates an additional feature, an additional threshold and nonlinearity, which could be useful for realizing some of the NMC algorithms.

In parallel with the scaling up the described superconductor ANN and memory, it would be interesting to start developing specific algorithms and software utilizing unique features of the bioSFQ networks.

In addition to scaling up, another essential improvement is the improvement of energy efficiency. This can be accomplished by reducing the critical currents of all Josephson junctions and by implementing the ERSFQ biasing technique [30] or the ultimate SFQ biasing technique suggested recently [31]. Unfortunately, the latter requires more advanced fabrication processes [32] and therefore extra fabrication time and expenses. We keep these improvements in mind for the prospective work.

ACKNOWLEDGMENT

The numerical simulations were performed using PSCAN2 software package developed by Pavel Shevchenko [33]. We thank Coenrad Fourie for assistance with InductEx software [34] used for inductance extraction from the circuit layouts. We are also grateful to Vladimir Bolkhovsky and Ravi Rastogi for overseeing the wafer fabrication.

This research was supported by the Under Secretary of Defense for Research and Engineering under Air Force Contract No. FA8702-15-D-0001. Any opinions, findings, conclusions or recommendations expressed in this material are those of the author(s) and do not necessarily reflect the views of the Under Secretary of Defense for Research and Engineering. Delivered to the U.S. Government with Unlimited Rights, as defined in DFARS Part 252.227-7013 or 7014 (Feb 2014). Notwithstanding any copyright notice, U.S. Government rights in this work are defined by DFARS 252.227-7013 or DFARS 252.227-7014 as detailed above.